# Formation of C1 oxygenates by Activation of Methane on B, N Co-doped Graphene Surface Decorated by Oxygen Pre-covered Ir13 Cluster: A First Principles Study

Jemal Yimer Damte[a*], Jiri Houska[a†]

[a] Department of Applied Physics, University of West Bohemia, Univerzitni 2732/8, 301 00 Plzen 3, Pilsen, Czech Republic

Email: * damtejem@ntis.zcu.cz, † jhouska@kfy.zcu.cz

# Abstract

We employ density functional theory (DFT) to investigate the adsorption and dehydrogenation of methane on the BNG-Ir13 cluster at both low and high oxygen coverage. The DFT calculations show that the low-oxygen-coverage BNG-Ir13 cluster (BNG-Ir13O cluster) forms methanol and formaldehyde with a lower activation energy barrier compared to the high-oxygen-coverage BNG-Ir13 cluster. Furthermore, the results reveal that the BNG-Ir13 cluster with low oxygen coverage has a higher methane adsorption energy and a lower activation energy barrier for methane dissociation compared to the high-oxygen-coverage BNG-Ir13 cluster. Quantitatively, the methane adsorption energy on the low-oxygen-coverage BNG-Ir13 cluster is -0.44 eV, and the second dehydrogenation of methane is the rate-determining step with an energy barrier of 1.24 eV, in both cases lower numbers than those observed for the high-oxygen-coverage BNG-Ir13 cluster. Moreover, at a controlled reaction temperature, $CH_3$ and $CH_2$ species were found to be the most abundant species on the oxygen-pre-covered BNG-Ir13 cluster, and C–O coupling reactions were considered. Hydrogen recombination was also observed, indicating that hydrogen molecules could be formed on this surface. Overall, the study suggests that the low-oxygen-coverage BNG-Ir13 cluster could be a promising catalyst for selectively converting methane to methanol, formaldehyde, and hydrogen production.

# 1. Introduction

Natural gas predominantly consists of methane, which is extensively utilized as a primary fuel. The utilization of methane has gained considerable interest in recent times due to the depletion of fossil fuel reserves and the requirement for innovative technologies. However, the emission of methane directly or via oxidation to $CO_2$, contributes to the issue of global warming. Therefore, methane conversion is an attractive alternative to maintaining environmental safety and producing valuable chemicals [1-4]. The prevalent approach for transforming methane into various value-added chemicals involves an indirect process known as synthesis gas production. The synthesized gas is subsequently utilized to produce essential commodities via the Fischer-Tropsch process. Even though synthesis gas production is a proven technology for methane conversions, it is the most cost-intensive and multi-step process [5-8]. Hence, considerable literature has been published to find the most attractive routes in converting methane to value-added products. The direct selective oxidation of methane to C1 oxygenates (methanol and formaldehyde) can be essential in addressing the cost issue by removing the synthesis gas production.

$$2CH_4 + O_2 \rightarrow 2CH_3OH, \Delta H_{298K} = -252.8 \text{ kJ/mol} \quad (1)$$

$$CH_4 + O_2 \rightarrow CH_2O + H_2O, \Delta H_{298K} = -282.6 \text{ kJ/mol} \quad (2)$$

The process of converting methane to methanol and formaldehyde (reactions 1 and 2) is thermodynamically favorable, making the oxidative transformation of methane with oxygen an attractive method for methane conversion. These products are commonly used as feedstock in the petrochemical industry and can be easily transported at a low cost. However, achieving selective oxidation of methane has been a topic of intense debate in the scientific community. The direct conversion of methane to C1 oxygenates is particularly challenging due to the high reactivity of the desired product, which often leads to further oxidation to carbon dioxide, resulting in reduced product selectivity.[9-14]

Over the last few decades, numerous studies have yielded significant insights into different catalysts for the oxidation of methane to C1 oxygenates. Transition metal oxides have shown promise as effective catalysts for this process. Both theoretical and experimental investigations have been conducted to explore their exceptional catalytic activity in selectively converting methane to valuable chemicals. [15-20]. For example, selective oxidation of methane to methanol has been reported on Fe/SiO$_2$ catalyst using oxygen as an oxidant at high-temperature conditions (700 °C). Due to the stability of methane, this catalyst operates at high temperatures to cleave the initial C−H bond of methane and produce free CH$_3$ radicals on the surface. Under such conditions, the CH$_3$ radical couples with surface oxygen to produce C1 oxygenates. However, this oxidant, at high-temperature conditions, facilitates the over-oxidation process and overwhelms the selective oxidation of methane to C1 oxygenates. Even though alternative oxidants like gaseous sulfur are used for the selective oxidation of methane and can minimize the over-oxidation process, the performance is deficient compared to oxygen. Therefore, there is a need for efficient catalysts for activating methane and its selective oxidation to C1 oxygenates [21-25].

Numerous studies revealed that metal clusters effectively catalyze reactions with attractive characteristics. Transition metal clusters are found to be efficient catalysts due to their importance in various industrial applications. Iridium clusters play significant roles and are used as catalysts for various hydrocarbon reactions for activating C−C, C−N, and C−H bonds. Low temperature activation of methane has been investigated on IrO$_2$ (110) surface[26]. Besides this, iridium clusters supported on metal oxide surfaces and doped graphene show an efficient catalytic activity for alkane dehydrogenation and promising catalysts for activating the C−H bond. Boron nitrogen co-doped graphene improves the catalytic activity of the catalyst, increases the thermo-stability of the graphene, prevents the clustering of metals, and enhances the dispersion of the iridium cluster [27-37]. Based on this, we theoretically investigate the methane activation and the C-O coupling reactions on boron nitrogen co-doped graphene decorated by oxygen pre-covered iridium cluster (BNGIr13O cluster). This work aims to convert methane to C1 oxygenates using an oxidant on boron nitrogen co-doped graphene decorated by iridium cluster under mild temperature conditions.

## 2. Computational Details

All calculations were performed in the framework of density functional theory (DFT) with the program package Vienna ab initio simulation package (VASP)[38]. The exchange-correlation energy was calculated within the generalized gradient approximation (GGA) of the Perdew–Burke–Ernzerhof (PBE) functional[39], and the electron core interaction was described by the projector augmented wave (PAW) method[40]. The Van der Waals correction proposed by Girmme DFT+D3 was applied to expand the description of the Perdew–Burke–Ernzerhof (PBE) functional and inter-molecular dispersion interactions that were corrected[41]. A plane wave basis set with a kinetic energy of 400 eV was used, and the reciprocal space was sampled with a 3 x 3 x 1 Monkhorst–Pack k-point grid[42]. The boron nitrogen co-doped graphene surface model representing a periodic 6 x 6 super-cell unit-cell with a 15 Å vacuum thickness was used to separate the surface from its periodic images along the z direction. All atoms were allowed to relax in optimized geometry calculations, and spin polarization was considered.

The adsorption energy of the intermediates was calculated using the following definition:

$$E_{ads} = E_{surface/molecule} - E_{surface} - E_{molecule} \qquad 1$$

Where $E_{surface/molecule}$ is the total energy of surface together with a molecule, $E_{surface}$ is the energy of a clean surface and $E_{molecule}$ is the calculated energy of a molecule. The Climbing Image Nudged Elastic Band (CI-NEB) method[43] was applied to calculate the transition state of various elementary steps involved in methane dissociation on boron nitrogen co-doped graphene decorated by Ir13 (BNG-Ir13) cluster. The transition state corresponds to the highest energy along the reaction coordinate defined by NEB calculation. Vibrational frequencies were calculated for the optimized geometry intermediates and transition state structure; furthermore, zero-point energy correction was included in the reaction energetics, which is calculated as follows:

$$ZPE = \sum \frac{1}{2} h\nu, \qquad 2$$

Where h is Planck's constant and ν is the vibration frequency. Possible optimized geometries of Ir13 cluster on boron nitrogen co-doped graphene surface were considered. The possible structures and the calculated energies are shown in Figure S1 and Table S1 in the supporting information, respectively. The most stable oxygen pre-covered BNG-Ir13 cluster is shown below in Figure 1.

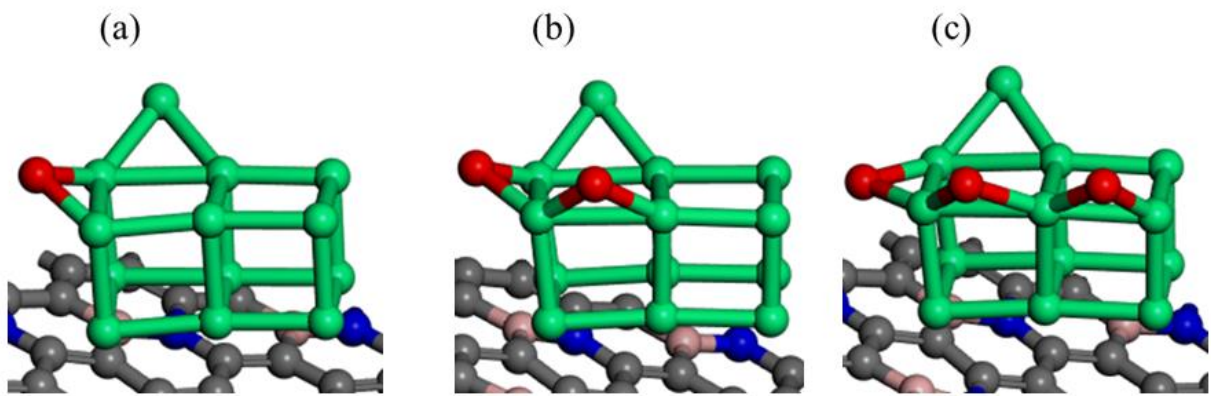

Figure 1. The most stable optimized structures of oxygen pre-covered Ir13 cluster decorated on boron nitrogen co-doped graphene surface (a) BNG-Ir13O cluster (b) BNG-Ir13O2 cluster and (c) BNG-Ir13O3 cluster, atomic spheres: green, Ir; gray, C; deep blue, N; pink B; red, O.

# 3. Results and discussion

## 3.1. Oxygen pre-covered BNG-Ir13 cluster

Oxygen is frequently used as an oxidant for oxidative coupling of methane. However, owing to the high energy barrier for dissociation of oxygen molecule on BNG-Ir13 cluster, in this study we have adsorbed and dissociated water in order to get oxygen atom. Various adsorption sites were analyzed for the adsorption of water onto the BNG-Ir13 cluster, revealing that water adsorbs perpendicularly to the surface, with the oxygen binding to the top site of Ir in the cluster. The most stable adsorption energy observed for water on this surface was -0.78 eV, which exceeds previously reported values [44-46]. The optimized geometry of the adsorption of water on BNG-Ir13 cluster is shown in the first panel of Figure S2 in the supporting information. Following the identification of stable water adsorption on the BNG-Ir13 cluster, we investigated the sequential dehydrogenation of water to generate oxygen atoms. The optimized geometry of the initial, transition, and final states of water dehydrogenation on the cluster are illustrated in Figure S2 of the supporting information. The initial dehydrogenation of adsorbed water is the most facile step and exothermic in nature (as shown in Table S2). However, the second dehydrogenation step, which involves dehydrogenating OH to O+H to produce oxygen atoms on the BNG-Ir13 cluster, is the rate-limiting step and endothermic with a higher activation energy barrier. Consequently, a disproportionate reaction, $2OH \rightarrow O + H_2O$, was considered on the BNG-Ir13 cluster, resulting in the formation of oxygen atoms with a lower activation energy barrier (as indicated in Table S2), albeit with slightly endothermic reaction energy. Upon introducing the oxygen atom to the BNG-Ir13 cluster, it was observed that the oxygen atom favored adsorption at the bridge site of Ir in the cluster (as depicted in Figure 1). During dehydrogenation, the hydrogen atom remained on the

BNG-Ir13 cluster, and we subsequently studied hydrogen recombination to form a hydrogen molecule. The result confirms that the formation of the hydrogen molecule is possible in oxygen pre-covered BNG-Ir13 cluster. The optimized structures of the initial states, the transition states, and the final states are presented in Figure S3 in the supporting information.

## 3.2. Adsorption of methane and its intermediates on oxygen pre-covered BNG-Ir13 cluster

To examine the impact of oxygen coverage on methane dehydrogenations and the potential for C-O coupling reactions, we investigated the low and high oxygen coverage on the BNG-Ir13 cluster. This involved studying the adsorption of intermediates by progressively introducing oxygen to the BNG-Ir13 cluster. We explored potential oxygen adsorption sites on the cluster and discovered that the bridge site was the most stable configuration. To obtain the optimized structure of the BNGIr13O cluster, we added an oxygen atom to the BNGIr-13 cluster. We then repeated this process to obtain the optimized geometry of the BNGIr13O2 and BNGIr13O3 clusters, respectively.

In order to determine the optimal adsorption structures of methane, we have investigated the most stable geometric structures and adsorption energies of methane by varying the oxygen coverage on BNG-Ir13 cluster. The most stable optimized structures and the adsorption energies of methane including the geometric parameters are shown in Figure 2 and Table 1, respectively. After considering possible configuration of methane in all surfaces, the top site of Ir in oxygen pre-covered BNG-Ir13 cluster has been found to be the favorable adsorption site of methane. Methane adsorption energies on the BNG-Ir13O, BNG-Ir13O2, and BNG-Ir13O3 clusters are -

0.44 eV, -0.39 eV, and -0.36 eV, respectively. Compared to other intermediates (Table 1), methane has a lower adsorption energy due to its saturated nature and the greater distance between the molecule and the surface. When adsorbed onto the BNG-Ir13O cluster, the hydrogen in the C-H bond of methane pointing towards the surface becomes elongated to 1.13Å, while on other surfaces such as BNG-Ir13O2 and BNG-Ir13O3, it elongates to 1.12Å, suggesting that dissociation of $CH_4$ into $CH_3$ and H is more likely. The result shows that methane's adsorption energy decreases as oxygen co-adsorption increases, which is in agreement with previous theoretical and experimental results[47-49].

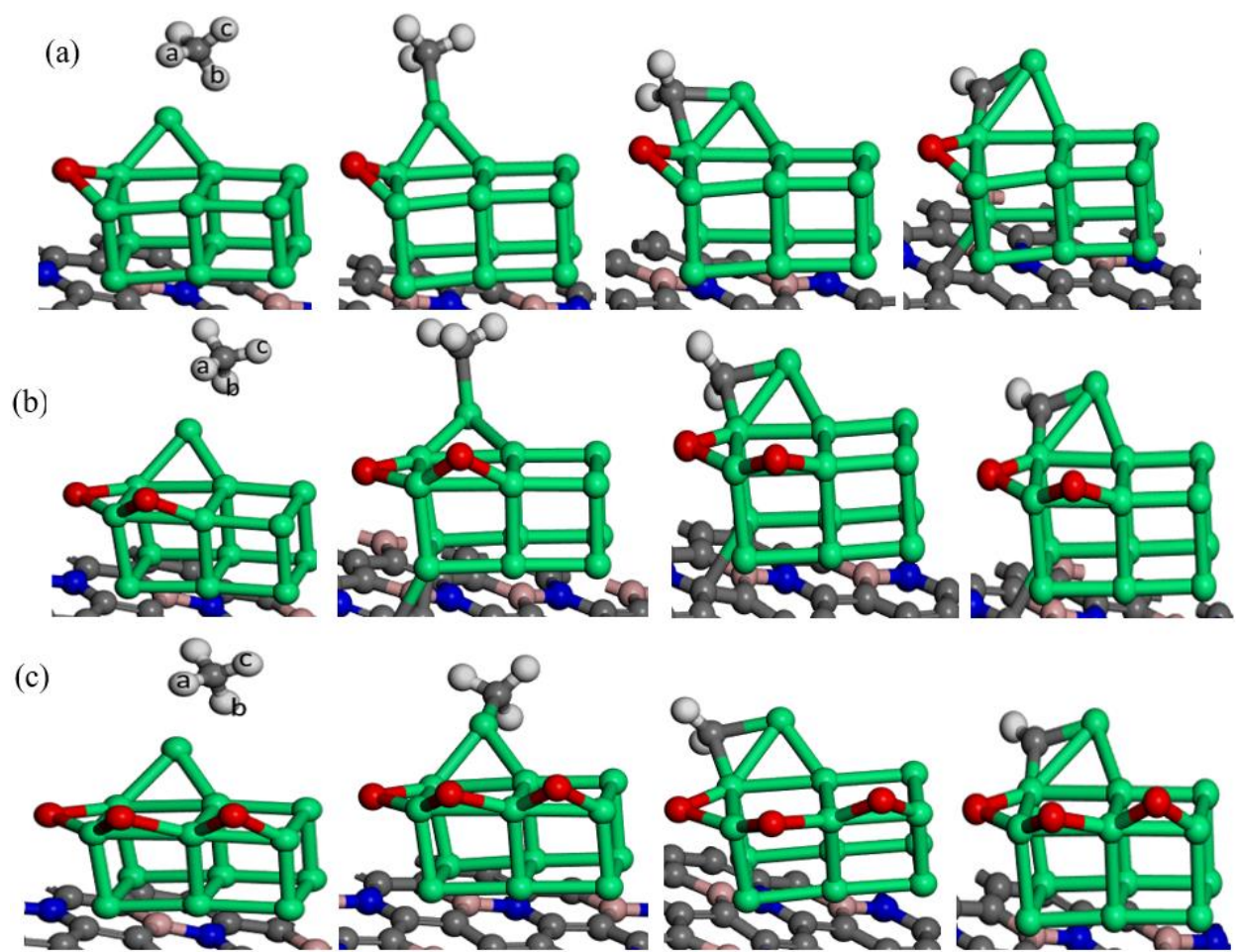

Figure 2. Optimized geometries of the most stable structures of intermediates on (a) BNG-Ir13O cluster, (b) BNG-Ir13O2 cluster and (c) BNG-Ir13O3 cluster, atomic spheres: green, Ir; gray, C; deep blue, N; pink, B; white, H; red, O.

The adsorption energy of methane is higher at low oxygen coverage of the BNG-Ir13 cluster and it decreases as oxygen co-adsorption increases due to the lateral adsorbate repulsion. The oxygen-pre-covered BNG-Ir13 cluster suppresses the adsorption energy of methane compared to methane adsorption on the BNG-Ir13 cluster (Table 1). However, when the oxygen coverage increases on the BNG-Ir13 cluster, the adsorption energy of unsaturated intermediates (contrary to that of saturated methane) increases.

Moreover, the bond length of Ir-C (2.45 Å) in the BNG-Ir13O cluster is shorter than that of the BNG-Ir13O2 cluster and BNG-Ir13O3 cluster, confirming that the strong interaction of methane and the surface and inconsistent with the higher adsorption energy of methane occurs on low oxygen coverage of BNG-Ir13 cluster than that of high oxygen coverage of BNG-Ir13 cluster. The result is in consistent with the electron density difference (EDD) plot of methane on the surface (Figure 3). It shows that more significant overlapping of orbitals on the BNG-Ir13O cluster occurred more than that of the BNG-Ir13O2 cluster and BNG-Ir13O3 cluster.

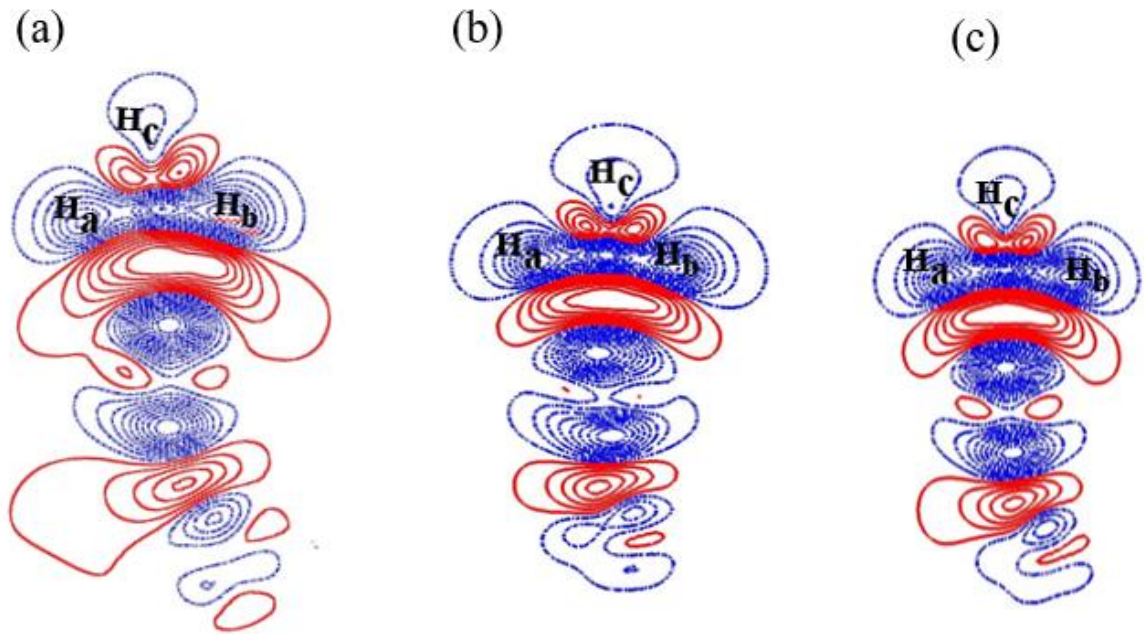

Figure 3 Electron density difference contour plots of $CH_4$ on (a) BNG-Ir13O cluster; (b) BNG-Ir13O2 cluster; and (c) BNG-Ir13O3 cluster, the solid red and dashed blue lines represent increasing and decreasing electron densities, respectively, Iso-surface value: 0.001

Adsorption of $CH_X$ intermediates plays a vital role in methane dehydrogenation, and the hollow site, bridge site, and top site have been considered. We found that Ir's top site and Ir's bridge site in the oxygen pre-covered BNG-Ir13 cluster are the most stable adsorption sites. The adsorption energy of $CH_X$ intermediates increases with decreasing number of hydrogen atoms (Table 1), which is in agreement with previous DFT studies [50-52]. Unsaturated intermediates such as methyl, methylene, and methyne are highly reactive and tend to form new bonds. Table 1 shows the chemisorption energies of CHx intermediates resulting from methane dissociation on an oxygen pre-covered BNG-Ir13 cluster, while Figure 2 illustrates the corresponding adsorption geometries.

Table 1 Adsorption energies ($E_{ads}$, eV), C–H bond lengths (Å), and Ir–C bond lengths (Å) of CHx (x = 0–4) on oxygen pre-covered BNG Ir13 cluster. The energy values in parentheses are calculated on BNG-Ir13 cluster.

| Surface | BNG-Ir13O | | |
|---|---|---|---|
| Species | $E_{ads}$ | d(C-$H_a$, $H_b$, $H_c$) | d(Ir-C) |
| $CH_4$ | -0.44 (-0.45) | 1.13, 1.12, 1.10 | 2.45 |
| $CH_3$ | -2.71 (-2.90) | 1.10 | 2.04 |
| $CH_2$ | -4.43 (-4.71) | 1.10 | 1.97, 2.05 |
| CH | -6.29 (-6.41) | 1.10 | 1.86, 1.89 |
| Surface | BNG-Ir13O2 | | |
| Species | $E_{ads}$ (eV) | d(C-$H_a$, $H_b$, $H_c$) | d(Ir-C) |
| $CH_4$ | -0.39 | 1.12, 1.12, 1.10 | 2.54 |
| $CH_3$ | -3.06 | 1.10 | 2.00 |
| $CH_2$ | -4.67 | 1.10 | 1.97, 2.05 |
| CH | -6.42 | 1.10 | 1.85, 1.91 |
| Surface | BNG-Ir13O3 | | |
| Species | $E_{ads}$ (eV) | d(C-$H_a$, $H_b$, $H_c$) | d(Ir-C) |
| $CH_4$ | -0.36 | 1.12, 1.12, 1.10 | 2.53 |
| $CH_3$ | -2.83 | 1.10 | 2.01 |
| $CH_2$ | -4.52 | 1.10 | 1.97, 2.06 |
| CH | -6.53 | 1.10 | 1.85, 1.90 |

## 3.3. Dissociation of methane on oxygen pre-covered BNG-Ir13 cluster

After the most stable adsorption geometries have been determined, methane dissociation occurs through a series of C-H bond cleavage steps, namely $CH_4$ to $CH_3$ and H, $CH_3$ to $CH_2$ and H, and $CH_2$ to CH and H. This study focuses on the dissociation of methane at varying oxygen coverage on the BNG-Ir13 cluster. The initial state is methane's stable adsorption site, and the final state is the co-adsorbed $CH_3$ and H. Table 2 provides the reaction energies and activation energies, while Figure 4 and Figure S4 displays the geometric optimization of the initial, transition, and final states. It is apparent that the hydrogen, directed towards the surface, abstracts and produces $CH_3$ and H species, which sit atop the Ir in an oxygen pre-covered BNG-Ir13 cluster. The first step of $CH_4$ dehydrogenation on the BNG-Ir13O cluster has a 0.24 eV activation energy barrier with an exothermic reaction energy of -0.31 eV. In comparison to the BNG-Ir13O cluster, the methane activation energy barrier increases to 0.39 eV on the BNG-Ir13O2 cluster, while the reaction energy remains exothermic at -0.44 eV. To examine the impact of oxygen, the oxygen coverage was increased, and methane's dehydrogenation was analyzed. On the BNG-Ir13O3 cluster, the initial dehydrogenation stage of methane requires an activation energy barrier of 0.45 eV and results in an exothermic reaction (-0.52 eV). Comparatively, the BNG-Ir13 cluster with high oxygen coverage displays lower methane adsorption energy and higher activation energy barriers for methane activation than the BNG-Ir13 cluster with low oxygen coverage.

During the second dehydrogenation step, the dissociation of methyl occurs to yield methylene and hydrogen. This process requires overcoming an activation energy barrier of 1.24 eV, 1.34 eV, and 1.43 eV on the BNG-Ir13O, BNG-Ir13O2, and BNG-Ir13O3 clusters, respectively. The resulting reaction energies are endothermic with values of 0.86 eV, 0.76 eV, and 0.90 eV on the BNG-Ir13O, BNG-Ir13O2, and BNG-Ir13O3 clusters, respectively. In this dehydrogenation step, the hydrogen is abstracted and resides on the top site of Ir in oxygen pre-covered BNG-Ir13 cluster. The methylene intermediate is then rotated to attain stable structures. The second dehydrogenation step is the step that controls the rate of reaction in both the BNG-Ir13O cluster and BNG-Ir13O2 cluster. In all surfaces, the activation energy barrier of the second dehydrogenation step has increased, causing the reaction energy to become more endothermic than the others, as shown in Table 2. In the oxygenated BNG-Ir13 cluster, the third dehydrogenation step proceeds by dehydrogenating $CH_2$ to CH and H. The activation energy barrier of this step is

1.03 eV, 1.33 eV, and 1.69 eV in the BNG-Ir13O cluster, BNG-Ir13O2 cluster, and BNG-Ir13O3 cluster, respectively. Although the reaction energy is slightly endothermic in the BNG-Ir13O cluster, it is more endothermic in the BNG-Ir13O2 and BNG-Ir13O3 clusters. Finally, the third dehydrogenation step controls the reaction rate in the BNG-Ir13O3 cluster.

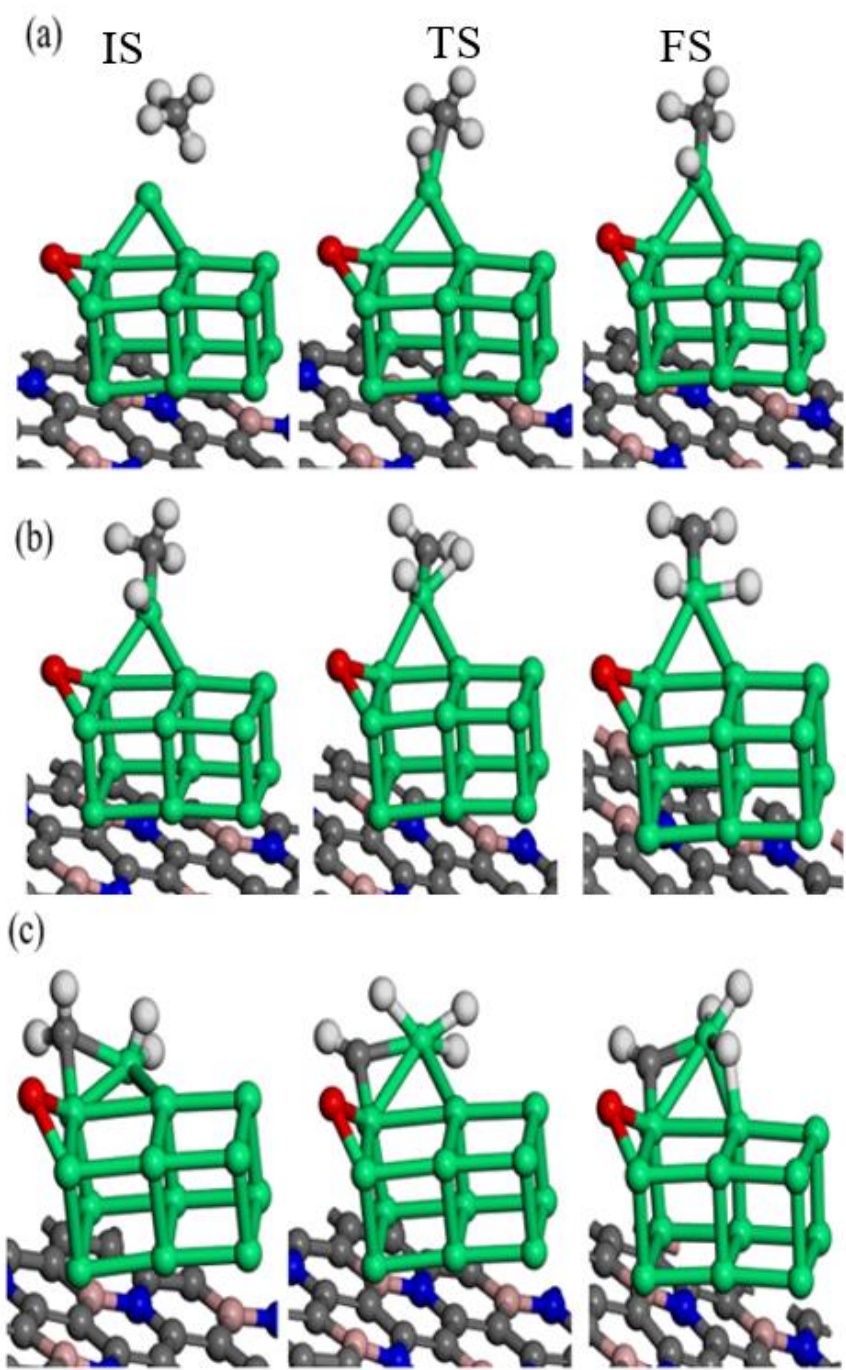

Figure 4. Initial states (IS), transition states (TS) and final states (FS) for dehydrogenation reaction of a) $CH_4$, b) $CH_3$, and c) $CH_2$ on oxygen pre-covered BNG-Ir13 cluster with low oxygen coverage (BNG-Ir13O cluster), atomic spheres: green, Ir; gray, C; deep blue, N; pink, B; white, H; red, O.

As the oxygen coverage increases in the BNG-Ir13 cluster, the activation energy barriers and reaction energies also increase, except for the first dehydrogenation step. Additionally, the presence of more pre-adsorbed oxygen on the BNG-Ir13 cluster reduces the interaction between methane and the surface, resulting in decreased catalytic activity compared to the BNG-Ir13O cluster. As presented in Table 2, the BNG-Ir13 cluster demonstrates high activity under oxygen-deficient conditions. Based on our results, the successive dehydrogenation reaction is as follows: $CH_3<CH_2<CH_4$ in the BNG-Ir13O and BNG-Ir13O2 cluster, whereas $CH_2<CH_3<CH_4$ in the BNG-Ir13O3 cluster. The chemisorbed oxygen with different coverages can alter the activation of methane, which agrees with the previous observations [53-55].

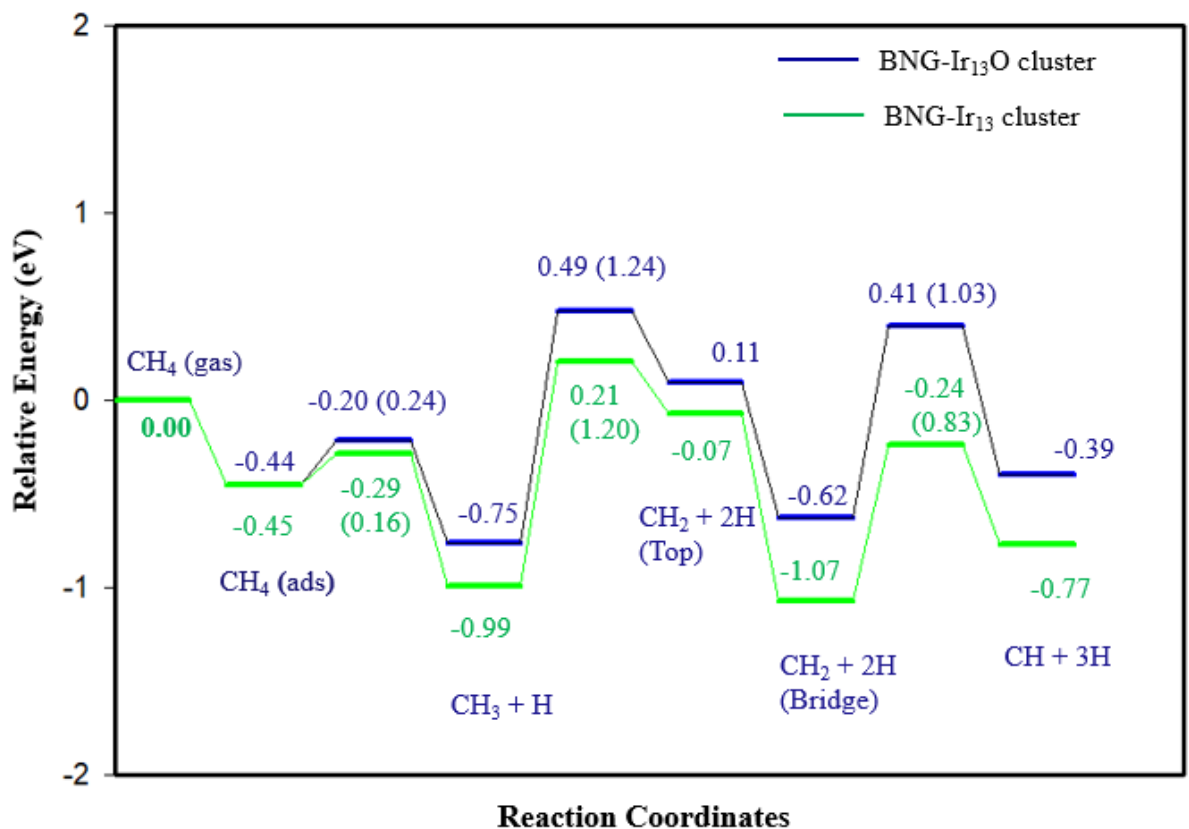

Figure 5. Potential energy diagram for dehydrogenation of CH$_4$ to CH and 3H on oxygen pre-covered BNG-Ir13 cluster and BNG-Ir13 cluster

Table 2 Activation energy barriers ($E_{act}$, eV), reaction energies ($\Delta E$, eV), activated C–H bond lengths of the transition state and imaginary frequencies (IMF, cm$^{-1}$) for dehydrogenation reactions in oxygen pre-covered BNG Ir13 cluster, and the energy values in parentheses are calculated on BNG-Ir13 cluster.

| Reactions | $E_{act}$ (eV) | $\Delta E$ (eV) | d (C–H) | IMF (cm$^{-1}$) |
|---|---|---|---|---|
| BNG-Ir13O cluster | | | | |
| CH$_4$ → CH$_3$ + H | 0.24 (0.16) | -0.31 (-0.54) | 1.43 | i783 |
| CH$_3$+H → CH$_2$ + 2H | 1.24 (1.20) | 0.86 (0.92) | 1.68 | i861 |
| CH$_2$+ 2H → CH + 3H | 1.03 (0.83) | 0.23 (0.30) | 1.67 | i760 |
| Reactions | $E_{act}$ (eV) | $\Delta E$ (eV) | d (C–H) | IMF (cm$^{-1}$) |
| BNG-Ir13O2 cluster | | | | |
| CH$_4$ → CH$_3$ + H | 0.39 | -0.44 | 1.46 | i855 |

| Reactions | $E_{act}$ (eV) | $\Delta E$ (eV) | d (C–H) | IMF (cm$^{-1}$) |
|---|---|---|---|---|
| CH$_3$ +H $\rightarrow$ CH$_2$+ 2H | 1.34 | 0.76 | 1.59 | i985 |
| CH$_2$ +2H $\rightarrow$ CH + 3H | 1.33 | 0.59 | 1.86 | i728 |
| BNG-Ir13O3 cluster | | | | |
| CH$_4$ $\rightarrow$ CH$_3$ + H | 0.45 | -0.52 | 1.51 | i821 |
| CH$_3$ + H $\rightarrow$ CH$_2$+ 2H | 1.43 | 0.90 | 1.62 | i931 |
| CH$_2$+2H $\rightarrow$ CH + 3H | 1.69 | 0.50 | 1.66 | i803 |

## 3.4. Conversion of Methane to C1 oxygenates

Figures 5 shows that the potential energy diagram of methane dehydrogenation reactions in oxygen pre-covered BNG-Ir13 cluster, where the dehydrogenation of CH$_3$ and CH$_2$ is kinetically and thermodynamically unfavorable. Thus, the C–O coupling reactions have been studied to form the C1 oxygenates by controlling the reaction temperatures and impeding further dehydrogenation of CH$_3$ and CH$_2$ species. Figure 6 and Figure S5 show the structures of initial states, transition states, and final states of the C–O coupling reactions in the oxygen-pre-covered BNG-Ir13 cluster. Table 3 and Table S3 list the activation energy barrier and reaction energy for the formation of methanol on the BNG-Ir13 cluster under low and high oxygen coverage conditions. The results indicate that the activation energy barrier for coupling CH$_3$ with a surface oxygen atom is lower in the BNG-Ir13O cluster with low oxygen coverage compared to the BNG-Ir13O2 and BNG-Ir13O3 clusters with high oxygen coverage (Table S3). When CH$_3$OH binds to the surface through the oxygen atom, the C-O axis tilts upward. The reaction energy is 0.48 eV, indicating an endothermic reaction, and the activation energy barrier is 1.37 eV. Hence, the formation of methanol is more feasible on an oxygen-pre-covered BNG-Ir13 cluster with low oxygen coverage than on the BNG-Ir13 cluster with high oxygen coverage.

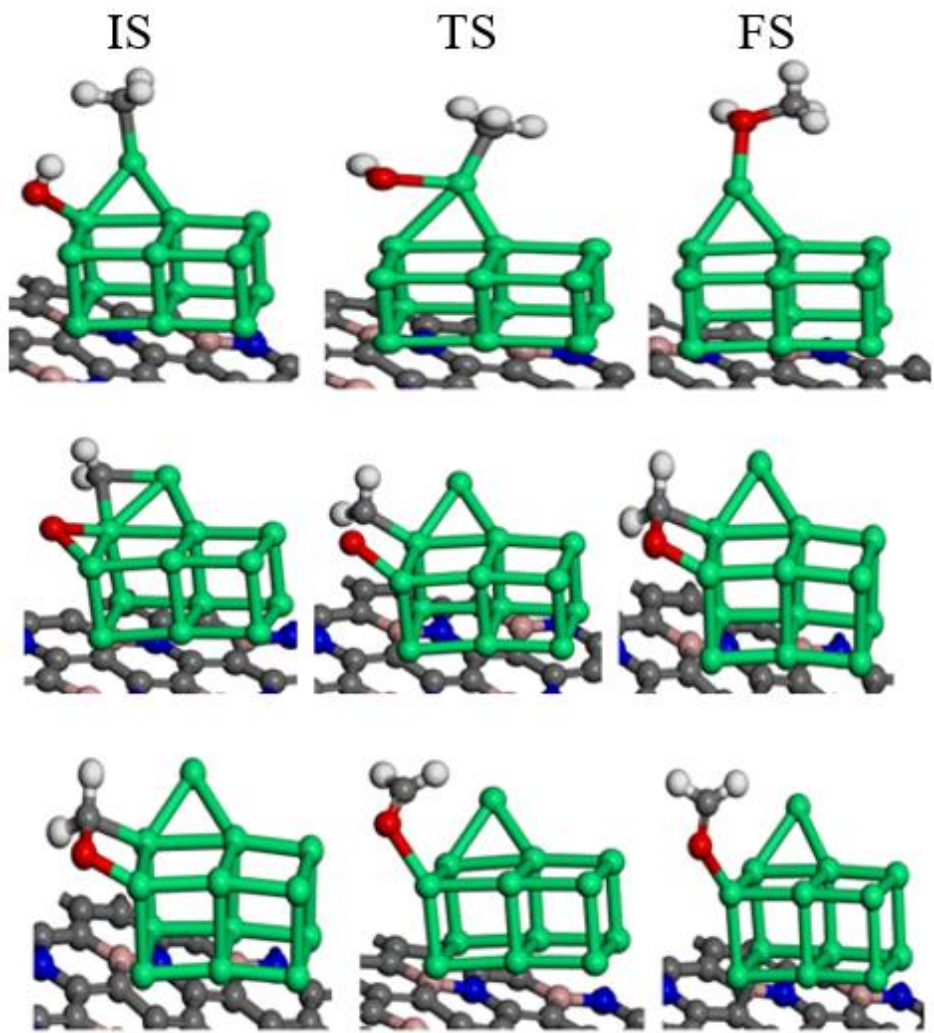

Figure 6. Optimized structures of initial states (IS), transition states (TS) and final states (FS) for C-O coupling reactions on oxygen pre-covered BNG-Ir13 cluster, atomic spheres: green, Ir; gray, C; deep blue, N; pink, B; white, H; red, O.

In addition, we examined the formation of formaldehyde on the BNG-Ir13 cluster under low and high oxygen coverage conditions. Our findings reveal that the activation energy barrier for formaldehyde formation is lower (1.30 eV) in the BNG-Ir13O cluster with low oxygen coverage. Moreover, the reaction energy (-0.32 eV) is exothermic, indicating favorable thermodynamic and kinetic conditions compared to the high oxygen coverage of BNG-Ir13 cluster

(Table S3). In the final part of our study, we investigated desorption of formaldehyde in the BNG-Ir13O cluster. Our results, presented in Figure 7, show that formaldehyde binds to the surface via the $\eta1$-$\eta1$(C, O) configuration, causing the Ir-C bond to break and transfer to the $\eta1$(O) configuration, which binds to the surface through the oxygen atom. The activation energy barrier for this process is 1.02 eV, while the reaction energy is 0.83 eV. The desorption energy of formaldehyde is 0.69 eV, which is lower than that of $CH_2$ dehydrogenation. The coupling reactions of $CH_3$ and $CH_2$ with surface oxygen atoms compete with $CH_3$ and $CH_2$ dehydrogenation reactions. Therefore, by controlling the reaction temperature, it is possible to form both methanol and formaldehyde at moderate temperature conditions in oxygen-pre-covered BNG-Ir13 clusters with low oxygen coverage.

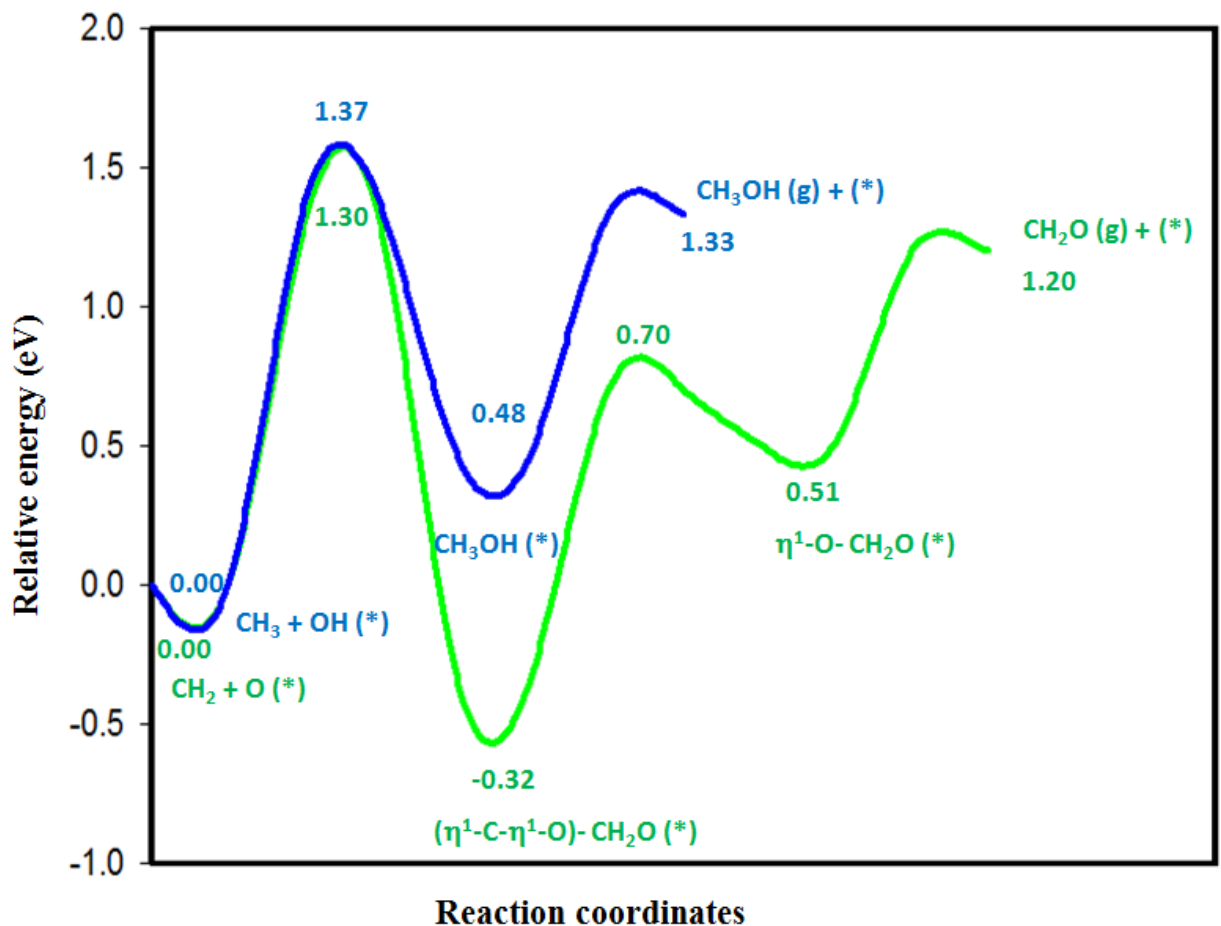

Figure 7. Potential energy diagram for methanol (blue color) and formaldehyde (light green) formation in low oxygen coverage of BNG-Ir13 cluster.

Table 3 Activation barriers ($E_{act}$, eV), reaction energies ($\Delta E$, eV) and imaginary frequencies (IMF, cm$^{-1}$) for C–O coupling reactions on oxygen pre-covered BNG Ir13 cluster.

| Reactions | $E_{act}$ (eV) | $\Delta E$ (eV) | IMF (cm$^{-1}$) |
|---|---|---|---|
| BNG-Ir13O | | | |
| $CH_3 + OH \longrightarrow CH_3OH$ | 1.37 | 0.48 | i544 |
| $CH_2 + O \longrightarrow CH_2O$ | 1.30 | -0.32 | i466 |
| $\eta 2\text{-(C-O) } CH_2O \longrightarrow \eta 1\text{-O } CH_2O$ | 1.02 | 0.83 | i381 |

## 4. Conclusions

Using density functional theory (DFT) methods, we examined the effect of varying oxygen coverage on the adsorption and activation of methane on the BNG-Ir13 cluster. Our results show that as the oxygen coverage increases, the adsorption energy of methane on the BNG-Ir13O cluster, BNG-Ir13O2 cluster, and BNG-Ir13O3 cluster slightly decreases, with values of -0.44 eV, -0.39 eV, and -0.36 eV, respectively. Additionally, we found that the low oxygen coverage of the BNG-

Ir13 cluster has a lower activation energy barrier for methane activation compared to the high oxygen coverage of the BNG-Ir13 cluster.

We identified that the second dehydrogenation step is the rate-determining step in oxygen-pre-covered BNG-Ir13 clusters (BNG-Ir13O and BNG-Ir13O2), while the third dehydrogenation step is the rate-determining step in BNG-Ir13O3. Furthermore, we observed that $CH_3$ and $CH_2$ dehydrogenations and coupling of $CH_3$ and $CH_2$ with surface oxygen atoms occur in competitive paths, resulting in these species being the most abundant in oxygen-pre-covered BNG-Ir13 clusters with both low and high oxygen coverage. We investigated C-O coupling reactions on all surfaces and found that the activation energy barrier to form methanol and formaldehyde is lower in the BNG-Ir13O cluster (1.37 eV and 1.30 eV, respectively) than in the BNG-Ir13O2 and BNG-Ir13O3 clusters. Therefore, the formation of methanol and formaldehyde is more favorable at low oxygen coverage of the BNG-Ir13 cluster under moderate temperature conditions.

Finally, we considered the recombination of hydrogen and found that it can be formed on this surface. Based on these findings, we propose that the low oxygen coverage of the BNG-Ir13 cluster has potential as a catalyst for the selective conversion of methane to methanol and formaldehyde and for hydrogen production.

## CRediT authorship contribution statement

**Jemal Yimer Damte:** Writing – review & editing, Writing – original draft, Investigation, Formal analysis, Data curation, Conceptualization. **Jiri Houska:** Writing – review & editing.

## Declaration of Interest Statement

The authors declare that they have no known competing financial interests or personal relationships that could have appeared to influence the work reported in this paper.

# Data availability

Data will be made available on request.

# Acknowledgement

This work supported by the Ministry of Science and Technology of Taiwan (MOST 105-2113-M-011-001). Furthermore, it was also supported by the project Quantum materials for applications in sustainable technologies (QM4ST), funded as project No. CZ.02.01.01/00/22_008/0004572 by Programme Johannes Amos Commenius, call Excellent Research.